\documentclass[a4paper,11pt]{article}
\pdfoutput=1 

\usepackage{jheppub} 

\usepackage[T1]{fontenc} 

\title{\boldmath Towards Quantum Turbulence in Finite Temperature Bose-Einstein Condensates}

\author[a]{Shanquan Lan,}
\author[b,c]{Yu Tian,}
\author[a,d]{Hongbao Zhang}

\affiliation[a]{Department of Physics, Beijing Normal University, Beijing 100875, China}
\affiliation[b]{School of Physics, University of Chinese Academy of Sciences, Beijing 100049, China}
\affiliation[c]{Shanghai Key Laboratory of High Temperature Superconductors, Shanghai 200444, China}
\affiliation[d]{Theoretische Natuurkunde, Vrije Universiteit Brussel, and The International Solvay
Institutes, Pleinlaan 2, B-1050 Brussels, Belgium}

\emailAdd{shanquanlan@mail.bnu.edu.cn}
\emailAdd{ytian@ucas.ac.cn}
\emailAdd{hzhang@vub.ac.be}

\abstract{Motivated by the various indications that holographic superfluid is BCS like at the standard quantization but BEC like at the alternative quantization, we have implemented the alternative quantization in the dynamical holographic superfluid for the first time. With this accomplishment, we further
initiate the detailed investigation of quantum turbulence in finite temperature BEC by a long time stable numerical simulation of bulk dynamics, which includes the two body decay of vortex number caused by vortex pair annihilation, the onset of superfluid turbulence signaled by Kolmogorov scaling law, and a direct energy cascade demonstrated by injecting energy to the turbulent superfluid. All of these results share the same patterns as the holographic superfluid at the standard quantization, thus suggest that these should be universal features for quantum turbulence at temperatures order of the critical temperature.}

\begin{document}
\maketitle
\flushbottom

\section{Introduction and Motivation}
\label{intro}
Turbulence is defined loosely as sort of temporally and spatially complex state of fluid motion. There are two types of turbulence. One is classical turbulence (CT) in normal fluid, where the eddies can have arbitrary circulation. The other is quantum turbulence (QT) in superfluid, where the involved vortices are instead quantized by quantum mechanics. Due to the presence of quantized vortices in QT, the hydrodynamical description of turbulence breaks down in superfluid. Thus some new approaches are developed in condensed matter physics to address QT, such as Gross-Pitaevskii equation (GPE) for zero temperature superfluid\cite{pitaevskii1961vortex,gross1961structure,gross1963hydrodynamics}, and Zaremba-Nikuni-Griffin (ZNG) formalism to take into account the finite temperature effect\cite{griffin2009bose}. However, both of them are dealing with the weakly interacting systems. Here comes holography as a complementary tool to attack QT at strongly coupled regime. Holographic duality provides us with a powerful framework, in which a complete description of a strongly coupled quantum many-body system, valid at all scales, can be encoded in a classical gravitational system with one extra dimension\cite{maldacena1998top,gubser1998gauge,witten1998anti}. In particular, the finite temperature effect can be beautifully implemented by putting a black hole in the dual bulk.

In order to address QT, one is required to first construct a holographic model for superfluid phase. Actually this has been successfully achieved by putting a U(1) gauge field minimally coupled to a complex scalar field with $m^2L^2=-2$ in AdS Schwarzschild, where $L$ is AdS curvature radius\cite{3h1,3h2}. As a result, when the temperature decreases to a certain critical value, the black hole will become hairy with the complex scalar field condensed, which corresponds exactly to the phase transition to a superfluid from a normal fluid on the boundary. Based on this simplest holographic superfluid model, the two dimensional QT at temperatures of order the critical temperature is investigated by holography in the seminal work \cite{chesler2013holographic}. It is found that although QT shares the same Kolmogorov energy spectrum as CT, QT exhibits a direct energy cascade from the large scales to the small scales, which is distinct from the inverse energy cascade in CT. Further work shows that the finite temperature leads to the decrease of vortex number, the behavior of which is well captured by vortex pair annihilation to sound waves\cite{Ewerz2014tua,du2014holographic}.

It is noteworthy that the above work on holographic superfluid turbulence focuses exclusively on the standard quantization\cite{chesler2015defect}. But as investigated from various aspects in \cite{Keranen2009ss,keranen2010inhomogeneous}, this standard quantization might give rise to a BCS-like superfluid while the alternative quantization might correspond to a BEC-like superfluid. For example, the density depletion in the core of both a single dark solition and vortex seems to be saturating at $40\%$ for the standard quantization and at $100\%$ for the alternative quantization when the temperature is lowered. This is reminiscent of a loosely bound BCS-like fermion pair condensed system and a more tightly bound BEC-like fermion pair condensed system, respectively. With this in mind, one is tempted to ask whether the aforementioned patterns for QT are universal, or dependent on whether the superfluid is BCS-like or BEC-like. As such, it is significant for one to explore what happens to QT for the alternative quantization. At first sight, the alternative quantization appears to be technically a routine generalization of the standard one by simply changing Dirichlet boundary condition to Neumann one at the AdS boundary. This is actually the case for the situation without the temporal evolution, where the Schwarzschild coordinates are preferred. But it turns out to be a non-trivial numerical challenge to implement the alternative quantization for the situation involving the temporal evolution like QT, where the infalling Eddington coordinates are favorably adopted. The main purpose of this paper is to conquer this numerical challenge and explore holographic superfluid turbulence at the alternative quantization, where the focus is put on the possible universal features such as the vortex number decay behavior, the scaling law, and energy cascade direction. Various aspects, especially the time evolution of vortex number and the scaling behavior in the kinetic energy spectrum, are investigated more carefully than in previous works.

The paper is organized as follows. In the next section, we shall develop the holographic superfluid model at the alternative quantization in the infalling Eddington coordinates and elaborate on the relevant numerics for our numerical simulation of the bulk dynamics. Then we present our numerical results on the decrease of vortex number by vortex pair annihilation, the scaling behaviors in the kinetic energy spectrum, and energy cascade direction separately in Section 3, Section 4 and Section 5. In the end, we summarize what we have founded with some discussions.

\section{Holographic Setup and Relevant Numerics}
\label{setup}
A simple holographic model for the two dimensional superfluid is a gravitational system in asymptotically AdS$_{4}$ spacetime, where dual to the conserved current $j^{a}$ and the condensate operator $O$ in the superfluid, a dynamical U(1) gauge field $A_{a}$ and a complex scalar field $\Psi$ with mass $m$ are introduced respectively and minimally coupled by the charge $q$. After rescaling the matter fields, the corresponding bulk action can be written as\cite{3h1,3h2}
\begin{equation}
S=\frac{1}{16\pi{G}}\int_{\mathcal{M}}\mathnormal{d}^{4}x\sqrt{-g}(R+\frac{6}{L^{2}}+\frac{1}{q^{2}}\mathcal{L}_{matter}).
\end{equation}
Here $G$ is the Newton's constant, and the matter Lagarangian reads
\begin{equation}
\mathcal{L}_{matter}=-\frac{1}{4}F_{ab}F^{ab}-|D\Psi|^{2}-m^{2}|\Psi|^{2},
\end{equation}
where $D=\nabla-iA$ with $\nabla$ the covariant derivative compatible to the metric. We are required to work in the regime with $\frac{L^{2}}{16\pi{G}}\gg1$ such that classical gravity is reliable, which corresponds to the large $N$ limit of the dual boundary system. A black hole solution corresponds to placing the boundary system at the Hawking temperature. The bulk gauge field $A_{t}=\mu$ at the AdS boundary corresponds to the chemical potential of the dual system. If the charge $q$ and conformal dimension
\begin{equation}\label{eqdelta}
\Delta=\frac{3}{2}\pm\sqrt{\frac{9}{4}+m^{2}L^{2}}
\end{equation}
of the scalar operator $O_{\pm}$ lie in certain range, taking the chemical potential sufficiently large will drive the bulk scalar field $\Psi$ to condense via Higgs mechanism. As a result, the black hole carries a scalar hair outside the horizon, which corresponds to a superfluid phase on the boundary.

The variation of action above will give rise to 16 equations of motion, including 10 from Einstein equation, 4 from Maxwell equation and 2 from Klein-Gordon equation. For simplicity, in this paper we shall ignore the backreaction of matter fields onto the metric, which can be implemented by taking the large $q$ limit. Thus the background geometry is simply required to be a solution to vacuum Einstein equation with a negative cosmological constant.  Since we are considering the superfluid at finite temperature, we take the Schwarzschild black brane solution as our background geometry, which can be expressed in the infalling Eddington coordinates as
\begin{equation}
ds^{2}=\frac{L^{2}}{z^{2}}(-f(z)dt^{2}-2dt dz+dx^{2}+dy^{2}),
\end{equation}
where the blackening factor $f(z)=1-(\frac{z}{z_{h}})^{3}$ with $z=z_{h}$ the horizon and $z=0$ the AdS boundary. As alluded to above, the dual boundary system is placed at Hawking temperature, which is given by
\begin{equation}
T=\frac{3}{4\pi{z_{h}}}.
\end{equation}
On top of this background geometry, the equations of motion for the matter fields can then be written as
\begin{equation}
D_{a}D^{a}\Psi-m^{2}\Psi=0,\nabla_{a}F^{ab}=i(\overline{\Psi}D^{b}\Psi-\Psi\overline{D^{b}\Psi}).
\end{equation}
In what follows, we will take the units in which $L=1,16\pi Gq^{2}=1$, and $z_{h}=1$, and work with the case of $m^{2}=-2$, in which the conformal dimension of the dual condensate operators $O_{\pm}$ can be $\Delta=2,1$, corresponding to the standard and alternative quantization respectively\cite{klebanov1999ads}. In the axial gauge $A_{z}=0$, the asymptotic solution of $A$ and $\Psi$ near the AdS boundary can be expanded as
\begin{equation}\label{asymp}
A_{\mu}=a_{\mu}+b_{\mu}z+o(z),\Psi=z[\phi+\psi{z}+o(z)].
\end{equation}
Then according to the holographic dictionary, the expectation value of $j^{\mu}$ and $O_{\pm}$ can be obtained explicitly as the variation of renormalized bulk on-shell action with respect to the sources. For the standard quantization, we have\cite{wjl}
\begin{eqnarray}
\langle j^{\mu}\rangle=\frac{\delta{S_{+}}}{\delta{a_{\mu}}}=\lim_{z\rightarrow 0}\sqrt{-g}F^{z\mu},
\end{eqnarray}
\begin{eqnarray}
\langle O_{+}\rangle=\frac{\delta{S_{+}}}{\delta{\phi}}=-\lim_{z\rightarrow 0}z\sqrt{-\gamma}(n_a\overline{D^a\Psi}+\overline{\Psi})=\overline{\psi}-\dot{\overline{\phi}}-ia_{t}\overline{\phi},
\end{eqnarray}
where the dot denotes the time derivative,and the renormalized action is given by
\begin{equation}
S_{+}=S-\int_{\mathcal{B}}\sqrt{-\gamma}|\Psi|^{2}
\end{equation}
with the counter term added to make the original action finite.

For the alternative quantization, one may naively think of $\psi$ as the source, with $\overline{\phi}$ the expectation value of the dual condensate operator. But this is not the case because we are working with the infalling Eddington coordinates rather than the Schwarzschild ones. Instead, it is $\tilde{\psi}=\langle \overline{O}_{+}\rangle$ that plays the role of the external source for the dual condensate operator. With this in mind, we thus have\cite{guo2016}
\begin{eqnarray}\label{current}
\langle j^{\mu}\rangle=\frac{\delta{S_{-}}}{\delta{a_{\mu}}}=\lim_{z\rightarrow 0}\sqrt{-g}F^{z\mu},
\end{eqnarray}
\begin{eqnarray}\label{operator}
\langle O_{-}\rangle=\frac{\delta{S_{-}}}{\delta\tilde{\psi}}=-\overline{\phi},
\end{eqnarray}
with
\begin{equation}
S_{-}=S+(\int_{\mathcal{B}}\sqrt{-\gamma}n_{a}\overline{D^{a}\Psi}\Psi+C.C.)+\int_{\mathcal{B}}\sqrt{-\gamma}|\Psi|^{2}.
\end{equation}
Although the resulting expectation value of condensate operator is still given by $\overline{\phi}$ up to an unimportant overall minus sign, it is compulsory to keep in mind that for the alternative quantization the spontaneously broken superfluid phase is accomplished via Higgs mechanism not with the boundary condition $\psi=0$, but with the source free boundary condition
\begin{eqnarray}\label{bcat}
\tilde{\psi}=\psi-\dot{\phi}+ia_{t}\phi=0
\end{eqnarray}
at the AdS boundary. It is this boundary condition that we shall employ to evolve the bulk scalar at the AdS boundary.

Vortices in superfluid are topological excitations with the quantized circulation. To see this, let us recall the definition of superfluid velocity
\begin{eqnarray}\label{velocity}
\boldsymbol{u}=\frac{\boldsymbol{j}}{|\phi|^{2}},\boldsymbol{j}=\frac{i}{2}[\overline{\phi}(\boldsymbol{\partial}-i\boldsymbol{a})\phi-\phi(\boldsymbol{\partial}+i\boldsymbol{a})\overline{\phi}].
\end{eqnarray}
With this, the winding number $\omega$ of a vortex is further defined as
\begin{equation}\label{omega}
\omega=\frac{1}{2\pi}\oint_{c}d\boldsymbol{x}\cdot\boldsymbol{u},
\end{equation}
where $c$ denotes a counterclockwise oriented path surrounding a single vortex. Note that close to the core of a single vortex with winding number $\omega$, the condensate $\phi\propto(\boldsymbol{z}-\boldsymbol{z_{0}})^{\omega}$ for $\omega>0$ and $\phi\propto(\boldsymbol{z}-\boldsymbol{z_{0}})^{-\omega}$ for $\omega<0$ with $\boldsymbol{z}$ the complex coordinate and $\boldsymbol{z_{0}}$ the location of the core. Whence the winding number must be an integer due to the single valued condensate. In addition, besides the vanishing magnitude of condensate at the core of a vortex, the corresponding phase shift around the vortex is given precisely by $2\pi\omega$. This characteristic property will be exploited as an efficient way to identify quantized vortices in our later vortex counting.

To address QT at the alternative quantization by holography, we would first like to impose the following boundary conditions onto the bulk fields, i.e.,
\begin{eqnarray}
a_{t}=\mu,\boldsymbol{a}=0
\end{eqnarray}
on top of (\ref{bcat}). In particular,  we set the chemical potential $\mu=2>\mu_{c}$ with $\mu_{c}\approx 1.12$ the critical chemical potential for the onset of a homogeneous and isotropic superfluid phase. Next we are required to prepare initial bulk configurations for the gauge field and complex scalar field at the Eddington time $t=0$. We shall investigate superfluid turbulence in an $100\times100$ periodic box, where the initial 32 vortex-antivortex pairs with the winding number $\omega=\pm 6$ are alternately placed at the square periodic lattice with lattice spacing $b=100/8$. As detailed in Appendix and illustrated in Fig.\ref{fig1}, it is not hard to construct an initial bulk configuration for the complex field $\Phi=\frac{\Psi}{z}$, dual to the aforementioned initial boundary state. Regarding the spatial components of gauge field $\boldsymbol{A}$, for simplicity but without loss of generality, we shall simply take $\boldsymbol{A}=0$ as the bulk initial configuration. Furthermore, the initial configuration for $A_{t}$ can be determined by the constraint equation
\begin{equation}\label{constraint}
\partial_{z}(\partial_{z}A_{t}-\partial\cdot\boldsymbol{A})=i(\overline{\Phi}\partial_{z}\Phi-\Phi\partial_{z}\overline{\Phi}),
\end{equation}
once we prescribe the second boundary condition $\partial_{z}A_{t}|_{z=0}=-\rho$ with $\rho$ the boundary charge density. As such, we take the initial charge density $\rho$ to be the value for the homogeneous and isotropic superfluid phase at the chemical potential $\mu$.

\begin{figure}
\begin{center}
\includegraphics[scale=0.4]{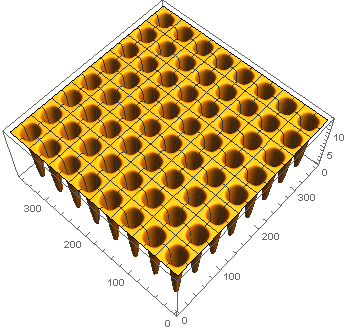}
\includegraphics[scale=0.4]{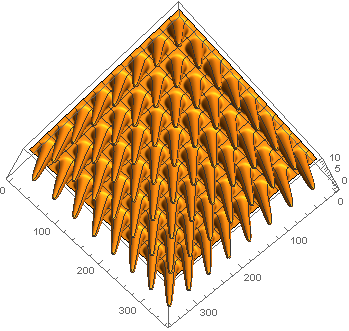}
\includegraphics[scale=0.3]{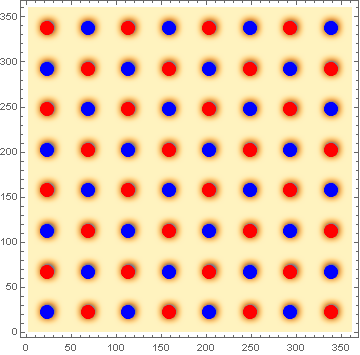}
\end{center}
\caption{The initial configuration of the turbulent superfluid at the chemical potential $\mu=2$. 32 vortex-antivortex pairs are alternatively placed in the periodic box. The first two panels are 3D plots of the condensate square $|\phi|^{2}$. The rightmost panel is the density plot of $|\phi|^{2}$ with the red points denoting vortices and the blue points denoting antivortices.}\label{fig1}
\end{figure}

With the above initial data and boundary conditions, the later time behavior of bulk fields can be obtained by the following evolution equations
\begin{eqnarray}\label{eqphi}
\partial_{t}\partial_{z}\Phi&=&iA_{t}\partial_{z}\Phi+\frac{1}{2}[i\partial_{z}A_{t}\Phi+f\partial^{2}_{z}\Phi+f'\partial_{z}\Phi\nonumber\\
&\,&+(\partial-iA)^{2}\Phi-z\Phi],
\end{eqnarray}
\begin{eqnarray}\label{eqa}
\partial_{t}\partial_{z}\boldsymbol{A}&=&\frac{1}{2}[\partial_{z}(\boldsymbol{\partial} A_{t}+f\partial_{z}\boldsymbol{A})+(\partial^{2}\boldsymbol{A}-\partial\boldsymbol{\partial}\cdot\boldsymbol{A})\nonumber\\
&\,&-i(\overline{\Phi}\partial\Phi-\Phi\partial\overline{\Phi})]-\boldsymbol{A}\overline{\Phi}\Phi,
\end{eqnarray}
\begin{eqnarray}\label{eqat}
\partial_{t}\partial_{z}A_{t}&=&\partial^{2}A_{t}+f\partial_{z}\boldsymbol{\partial}\cdot\boldsymbol{A}-\partial_{t}\boldsymbol{\partial}\cdot\boldsymbol{A}-2A_{t}\overline{\Phi}\Phi\nonumber\\
&\,&+if(\overline{\Phi}\partial_{z}\Phi-\Phi\partial_{z}\overline{\Phi})-i(\overline{\Phi}\partial_{t}\Phi-\Phi\partial_{t}\overline{\Phi}).
\end{eqnarray}
But in practice, to make our numerical simulation more reliable after a long time evolution, we adopt an alternative numerical scheme. Although we still evolve the scalar field $\Phi$ and the spatial components of gauge field $\boldsymbol{A}$ in time by (\ref{eqphi}) and (\ref{eqa}), we no longer use (\ref{eqat}) to achieve the time evolution of $A_{t}$ except at $z=0$, where (\ref{eqat}) reduces to the conservation law for charge density as follows
\begin{eqnarray}
\partial_{t}\rho=\partial_{z}\boldsymbol{\partial}\cdot\boldsymbol{A}|_{z=0}.
\end{eqnarray}
Then we can keep using the constraint equation (\ref{constraint}) to solve $A_{t}$ from $\Phi$ and $\boldsymbol{A}$ at every step of the later time evolution.

We numerically solve these non-trivial equations by employing the pseudo-spectral method with 28 Chebyshev modes in the $z$ direction and 361 Fourier modes in the $\boldsymbol{x}$ direction, as well as the fourth order Runge-Kutta method in time direction with the time step $\Delta t=0.05$. We evolve the system for a total period of time $t=450$, because it is expected that the intrinsic dynamics keeps unchanged without new interesting phenomenon popping up at later times. Finally, the vortex dynamics can be extracted from the near boundary behavior of bulk fields (\ref{asymp}) by the holographic dictionary (\ref{current}) and (\ref{operator}). Furthermore, we locate vortices by calculating the winding number (\ref{omega}) at each Fourier collocation point, which can be implemented by computing the total phase difference of $\phi$ around each plaquette formed by the nearest neighboring collocation points.

\section{Vortex Pair Annihilation and Decrease of Vortex Number}

\begin{figure}
\begin{center}
\includegraphics[scale=0.4]{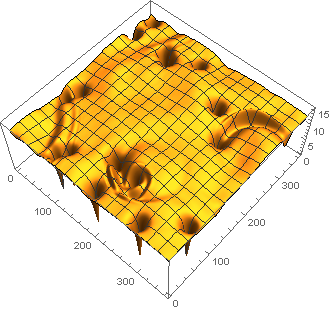}
\includegraphics[scale=0.4]{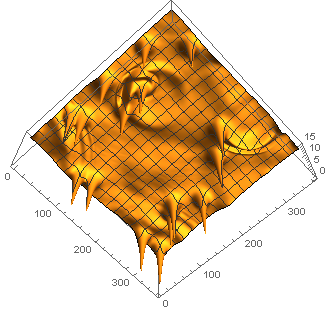}
\includegraphics[scale=0.3]{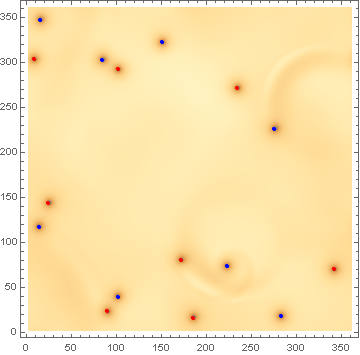}
\end{center}
\caption{The later time ($t=400$) configuration of the turbulent superfluid. The first two panels are 3D plots of the condensate square $|\phi|^{2}$. The rightmost panel is the density plot of $|\phi|^{2}$ with the red points denoting vortices and the blue points denoting antivortices. The ripples represent shock waves, and the bending structures represent solitons.}\label{fig2}
\end{figure}

We have obtained $18$ groups of data for the temporal evolution of superfluid turbulence at the chemical potential $\mu=2$. For the purpose of demonstration, we plot the configuration of turbulent superfluid at $t=400$ in Fig.\ref{fig2}. The typical behavior can be described as follows. Initially, the lattice of vortices with winding number $\pm6$ decays directly to the vortices with winding number $\pm1$ in the absence of vortices with intermediate winding numbers. Meanwhile, due to the repulsive force between the vortices with like-winding number, not only do the resulting clusters of six $\omega=\pm1$ vortices rotate around the cluster centers but also expand away from them. After a very short period, such an expansion makes the clusters of $\omega=1$ vortices meet the clusters of $\omega=-1$ vortices, which triggers the vortex-antivortex pair annihilation. The coalescence of vortex and antivortex cores leads universally to the formation of a crescent-shaped depletion of condensate, which then converts into a shock wave, dissipated gradually in the superfluid. Thus the vortex number gets decreased. This cluster collision induced vortex pair annihilation proceeds till the neighboring clusters are well mixed. Then it takes some time for these well mixed clusters to get dissolved before the superfluid is full of randomly distributed vortices. During this cross-over process, the vortex pair annihilation rate increases gradually. When the vortices get randomly distributed, the vortex pair annihilation is driven essentially by the thermal effect, thus will be absent at zero temperature. At this moment, the vortex pair annihilation rate should also arrive at its maximal value and will keep constant till the end of our evolution.

\begin{figure}
\begin{center}
\includegraphics[scale=0.8]{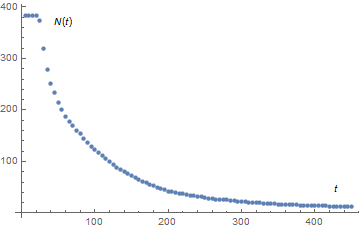}
\end{center}
\caption{The temporal evolution of averaged vortex number $N(t)$ in the turbulent superfluid over 18 groups of data  at the chemical potential $\mu=2$.}\label{fig3}
\end{figure}
\begin{figure}
\begin{center}
\includegraphics[scale=0.5]{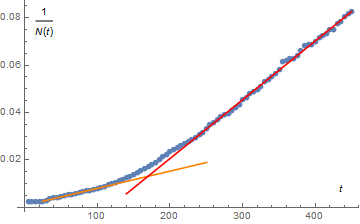}
\end{center}
\caption{The temporal evolution of inverse of averaged vortex number in the turbulent superfluid at $\mu=2$.}\label{fig4}
\end{figure}

To be quantitative, we would like to plot the temporal evolution of averaged vortex number over 18 groups of data in Fig.\ref{fig3}. As we see, the vortex number keeps unchanged when $t<25$, and then gets decreased with time. Since the decrease of vortex number is induced completely by vortex pair annihilation, we expect to have
\begin{equation}\label{ann}
\frac{dN(t)}{dt}=-\Gamma(t) N(t)^{2},
\end{equation}
where the vortex pair annihilation rate $\Gamma(t)$, as alluded to above, is supposed to be time dependent. To see this explicitly, we plot the time variation of inverse of averaged vortex number in Fig.\ref{fig4}. Note that (\ref{ann}) leads to
\begin{equation}
\frac{d\frac{1}{N(t)}}{dt}=\Gamma(t).
\end{equation}
Thus Fig.\ref{fig4} tells us that the vortex pair annihilation rate keeps invariant in the $25<t<110$ region, which corresponds to the aforementioned cluster collision induced vortex pair annihilation. In the cross-over region, namely $110<t<240$, the vortex pair annihilation rate increases gradually, which is followed by the purely thermal fluctuation driven vortex pair annihilation with a constant rate all the way to the end of our evolution.

\section{Kolmogorov Scaling Law and Onset of Quantum Turbulence}

Kolmogorov's $-5/3$ scaling law in the kinetic energy spectrum is derived by dimensional analysis under the assumption that there is an inertial range during which the energy is passed from one side to another without substantial loss. Thus not only is it supposed to be obeyed by CT but also by QT. Actually this scaling law can be used as a characteristic signal for the onset of turbulence. To this end, we define the kinetic energy density as
\begin{eqnarray}
\varepsilon_{kin}(t,\boldsymbol{x})\equiv\frac{1}{2}\overline{\boldsymbol{\nu}}(t,\boldsymbol{x})\cdot\boldsymbol{\nu}(t,\boldsymbol{x}),
\end{eqnarray}
where $\boldsymbol{\nu}=-\overline{\phi}\boldsymbol{u}$. By a spatial Fourier transform, the total kinetic energy can be written as an integral over momentum, i.e.,
\begin{eqnarray}
E_{kin}(t)=\int d^{2}x\varepsilon_{kin}(t,\boldsymbol{x})=\int_{0}^{\infty}dk\epsilon_{kin}(t,k),
\end{eqnarray}
where
\begin{eqnarray}
\epsilon_{kin}(t,k)=\frac{1}{2}\int_{0}^{2\pi}d\theta k\overline{\boldsymbol{\nu}}(t,\boldsymbol{k})\cdot\boldsymbol{\nu}(t,\boldsymbol{k})
\end{eqnarray}
with $(k,\theta)$ the polar coordinates for the momentum $\boldsymbol{k}$ space.
\begin{figure}
\begin{center}
\includegraphics[scale=0.5]{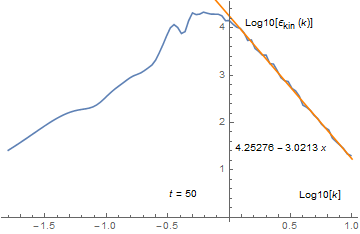}
\includegraphics[scale=0.5]{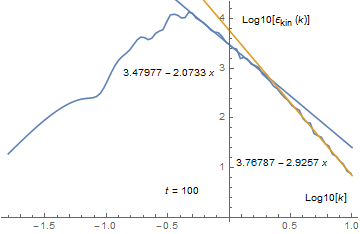}
\includegraphics[scale=0.5]{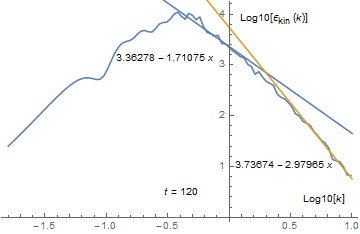}
\includegraphics[scale=0.5]{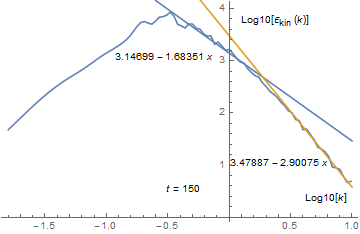}
\includegraphics[scale=0.5]{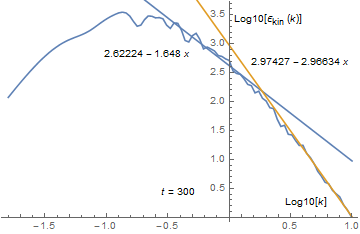}
\includegraphics[scale=0.5]{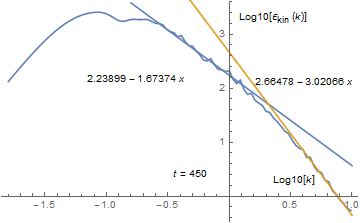}
\end{center}
\caption{The kinetic energy spectrum $\epsilon_{kin}(t,k)$ at time $t=50,100,120,150,300,450$ for one of the 18 groups data at chemical potential $\mu=2$. The horizontal axis denotes $log_{10}k$ and the vertical axis denotes $log_{10}\epsilon_{kin}(t,k)$. The blue and orange lines are fitting results of the spectrum.\label{fig5}}
\end{figure}

We plot the energy spectrum $\epsilon_{kin}(t,k)$ at the different times for one of 18 groups of data in Fig.\ref{fig5}, where the horizontal axis is $Log_{10}k$ and the vertical axis is $Log_{10}\epsilon_{kin}(t,k)$\footnote{The similar behaviors are expected to apply to other groups of data.}. As shown in Fig.\ref{fig5}, Kolmogorov scaling $k^{-5/3}$ shows up in the cross-over region and persists till the end of our simulation. This implies that only in the cross-over regime does the system get driven to a turbulent state, which is consistent with our intuition that the early alternately distributed vortices should not be regarded as a turbulent state. We also list the inertial range for Kolmogorov scaling law and averaged vortex spacing at different times in Table \ref{tab1}. As we see, the averaged vortex spacing falls well within the corresponding inertial region, which implies that Kolmogorov scaling law should be accounted for by the collective dynamics of turbulent vortices.
\begin{table}
\begin{center}
\begin{tabular}{|c||c|c|c|}
  \hline
  Time&Inertial Range in $\boldsymbol{k}$ Space \& $\boldsymbol{x}$ Space&Averaged Vortex Spacing\\
  \hline
  150 & 0.33-1.74, \ 3.6-19.0 & 11.8 \\
  \hline
  200 & 0.28-1.74, \ 3.6-22.4  & 15.35 \\
  \hline
  300 & 0.2-1.66, \ 3.8-31.4 & 21.32 \\
  \hline
  450 & 0.16-1.45, \ 4.3-39.3  & 28.86 \\
  \hline
\end{tabular}
\end{center}
\caption{Inertial range and averaged vortex spacing at different times, where the averaged vortex spacing is estimated by the formula $\sqrt{\frac{100\times100}{N(t)}}$.\label{tab1}}
\end{table}

On the other hand, besides Kolmogorov scaling law, we also see $k^{-3}$ scaling behavior, which appears after the
initial vortices of winding number $\pm 6$ finish decaying into the vortices of winding number $\pm 1$, and lasts all the way to the end of our evolution. This scaling has nothing to do with QT, but arises from the global rotation of vortices around the cluster centers at early times as well as the local rotation of superfluid around the vortex cores at all times.  Note that the IR end of this scaling region corresponds to the scale for the above regular rotation. As time passes, this scale moves from the cluster size $k\approx1$, asymptotically to the vortex size $k\approx3$,  signaling the transition from the dynamics dominated by the initial clusters of vortices at early times to the late state featured by randomly moving vortices.

\section{Energy Injection and Direct Energy Cascade}

\begin{figure}
\begin{center}
\includegraphics[scale=0.45]{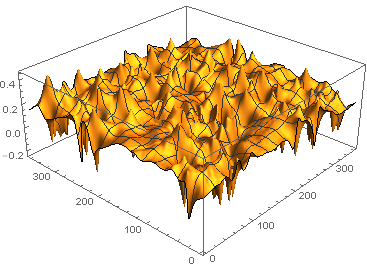}
\includegraphics[scale=0.4]{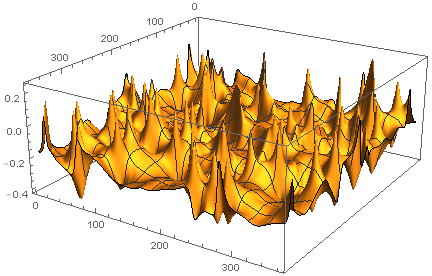}
\end{center}
\caption{The space distribution of injected power into the superfluid at $t=170$ (left panel) and $t=190$ (right panel).\label{fig7}}
\end{figure}

To see the energy cascade direction for our superfluid, we shall inject energy into the system at the IR and UV scales respectively and see how the energy flows in the kinetic energy spectrum. In particular, we would like to inject energy by switching on the electric field on the boundary. Note that the injected power is given by\cite{chesler2013holographic,tian2013poor,tian2014entropy}
\begin{eqnarray}\label{qby}
P&=&\int d^{2}xE_{i}(t,\boldsymbol{x})\langle j^{i}(t,\boldsymbol{x})\rangle|_{boundary}\nonumber\\
&=&\int d^{2}x\sum^{y}_{i=x}(\partial_{t}A_{i}-\partial_{i}A_{t})[(\partial_{t}A_{i}-\partial_{i}A_{t})-\partial_{z}A_{i}]|_{boundary}.
\end{eqnarray}
Thus we first choose the external source to be
\begin{eqnarray}
a_{x}(t)=\eta g(t-t_{0}), \,\,\,\, a_{y}(t)=-\eta g(t-t_{0}),
\end{eqnarray}
where $\eta$ is a small constant and $t_{0}=180$ and $g(t)$ is a gaussian function with width $8$. We turn on the sources at $t=150$ and turn them off at $t=210$. Although the driven sources are not functions of $x$ and $y$, the injected energy is not restricted at $k=0$, instead distributed mainly at low $k$s, and even a little bit at high $k$s\footnote{It is noteworthy that our energy cascade analysis is expected not to be affected by this tiny distribution of the injected energy at high $k$s.}. This is because there is a $-\partial_zA_i$ term in the current of Eq.(\ref{qby}), which depends on $x$ and $y$ in general. Also due to this term, the injected power is not positive everywhere. As a demonstration, we plot the space distribution of injected power at $t=170$ and $t=190$ in Fig.\ref{fig7}. As we see, not only is the injected power inhomogeneous, but also positive at some places and negative at other places. Actually before $t_0$ the total injected power is positive while after $t_0$ the total injected power is negative. For example, the total power at $t=170$ is $1928$ while it is $-1583$ at $t=190$.  As a result, there is a net positive energy injected into the system by the electric field.

\begin{figure}
\begin{center}
\includegraphics[scale=0.5]{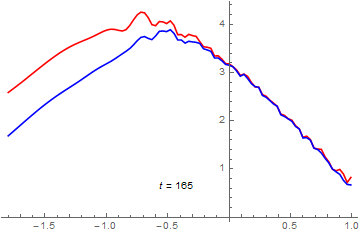}
\includegraphics[scale=0.5]{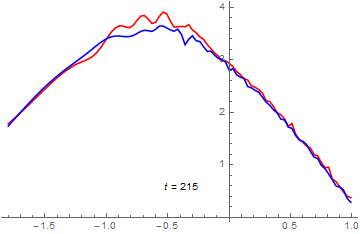}
\includegraphics[scale=0.5]{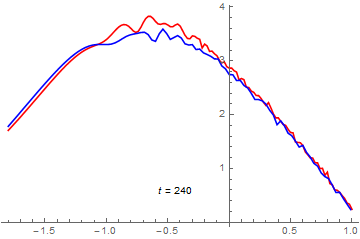}
\includegraphics[scale=0.5]{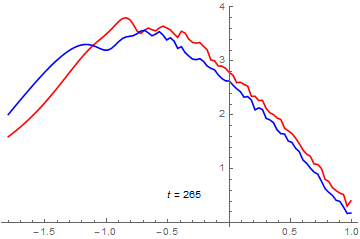}
\end{center}
\caption{The comparison of the kinetic energy spectra $\epsilon_{kin}(t,k)$ at time $t=165,215,240,265$ between the driven and undriven systems. The external sources are turned on at $t=150$ and turned off at $t=210$. The blue curve is the kinetic energy spectrum for the undriven superfluid while the red one is that for the driven superfluid. The injected energy at the IR is propagating to UV. \label{fig6}}
\end{figure}

Now we can see how the energy flows by comparing the evolution of the kinetic energy spectra between the driven and undriven systems with the same initial conditions, which is depicted in Fig.\ref{fig6}. As we expect, there is a positive energy injected to the turbulent superfluid at large scales, which then propagates deep into small scales and eventually results in an overall downwards shift of the energy spectrum at large scales and an overall upwards shift of the energy spectrum at small scales. This is a smoking gun for a direct energy cascade from the IR to the UV.

\begin{figure}
\begin{center}
\includegraphics[scale=0.5]{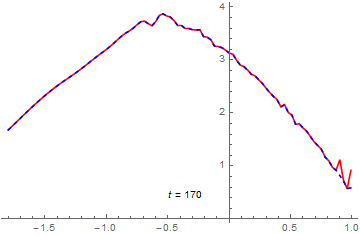}
\includegraphics[scale=0.5]{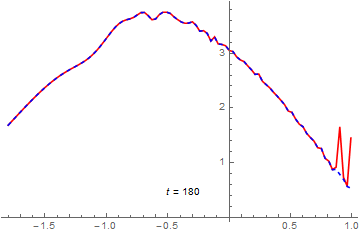}
\includegraphics[scale=0.5]{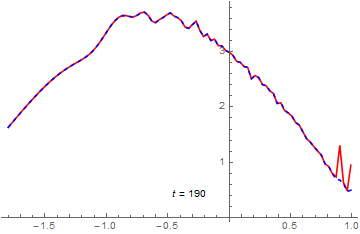}
\includegraphics[scale=0.5]{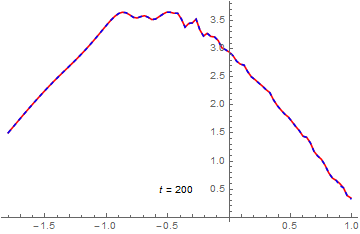}
\end{center}
\caption{The comparison of the kinetic energy spectra $\epsilon_{kin}(t,k)$ at time $t=170,180,190,200$ for the driven and undriven systems. The external sources are turned on at $t=150$ and turned off at $t=210$. The blue curve is the kinetic energy spectrum for the undriven superfluid while the red one is that for the driven superfluid. The injected energy at the UV is dissipated away rapidly, having no effect onto the kinetic energy spectrum at the IR.\label{fig8}}
\end{figure}

But nevertheless, to be reassuring, we also inject a net positive energy into the turbulent superfluid at high $k$s by the following external sources
\begin{eqnarray}
a_{x}(t)&=&\eta g(t-t_{0})(cos\frac{250\pi}{100}x+cos\frac{320\pi}{100}x), \nonumber\\
a_{y}(t)&=&-\eta g(t-t_{0})(cos\frac{250\pi}{100}y+cos\frac{320\pi}{100}y).
\end{eqnarray}
As depicted in Fig.\ref{fig8}, the injected energy at small scales dissipates away rapidly without affecting the kinetic energy spectrum at the IR. This is conceivable because the dominant dissipation mechanisms are expected to be vortex drag and vortex annihilation, both of which occur at small scales\cite{chesler2013holographic}.

\section{Summary and Discussion}

We have explored QT in finite temperature Bose-Einstein condensates by working with the alternative quantization in the simplest holographic model for the superfluid. By successfully accomplishing the long time stable numerical simulation of bulk dynamics in the infalling Eddington coordinates, we first investigate the temporal evolution of vortex number after shaking the lattice of vortices with winding number $\pm6$. As a result, the vortex
number keeps unchanged during the decay into the vortices of winding number $\pm1$, and then decreases due to the cluster collision induced vortex pair annihilation, which is further followed by a crossover to thermal fluctuation driven vortex pair annihilation. By fitting the data with the $N^2$ decay formula, we find that the vortex pair annihilation rate keeps constant in the cluster collision induced vortex pair annihilation, increases in the crossover process, and keeps a constant again in the thermal fluctuation driven vortex pair annihilation. This process is also exhibited somehow by the right movement of the IR end of the region for the scaling $k^{-3}$ in the kinetic energy spectrum. We further see Kolmogorov scaling law appearing in the crossover process, which can be regarded as the benchmark for the occurrence of the superfluid turbulence. At last we show that our superfluid has a direct energy cascade by injecting a positive energy into the system at both large and small scales.

Remarkably these results share the same features with those obtained in \cite{chesler2013holographic,Ewerz2014tua,du2014holographic} for the standard quantization case, although it is suggested in \cite{Keranen2009ss,keranen2010inhomogeneous} that the former resembles BEC while the latter resembles BCS. Consequently, it is natural to conjecture that the $N^2$ decay of vortex number, Kolmogorov scaling law and direct energy cascade should be universal patterns for the two dimensional superfluid turbulence at least at temperatures order of the critical temperature. On the other hand, it follows from the effective description of superfluid turbulence that QT at very low temperatures should have an inverse energy cascade with the $N^{5/3}$ decay behavior for the vortex number\cite{chesler2014vortex}. So it is intriguing to see how such a transition occurs when the temperature is lowered. But in order to achieve this by holography, one is required to take into account the backreaction of matter fields onto the metric. This is highly non-trivial and goes beyond the scope of this paper. We hope to address this issue in the near future and report our result somewhere else.

\acknowledgments

We are grateful to Yiqiang Du and Wenbiao Liu for their helpful discussions on the energy cascade program. S.L. is partially supported by NSFC with Grant Nos. 11235003, 11175019 and 11178007, as well as by "the Fundamental Research Funds for the Central Universities" with Grant
No.2015NT16. Y.T. is partially supported by NSFC with Grant No.11475179 and by the Opening Project of Shanghai Key Laboratory of High Temperature Superconductors(14DZ2260700). H.Z. is supported in part by the Belgian Federal Science Policy Office through the Interuniversity Attraction Pole P7/37, by FWO-Vlaanderen through the project G020714N, and by the Vrije Universiteit Brussel through the Strategic Research Program "High-Energy Physics". He is also an individual FWO Fellow supported by 12G3515N.

\section*{ Appendix: Construction of initial bulk configuration for $\Phi$ }

\begin{figure}
\begin{center}
\includegraphics[scale=0.5]{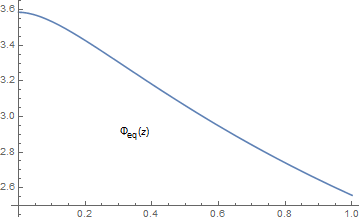}
\end{center}
\caption{The configurations of $\Phi_{eq}(z)$ at the chemical potential $\mu=2$.\label{fig9}}
\end{figure}

The purpose of this Appendix is to elaborate on how to construct the initial bulk configuration for $\Phi$ dual to the prescribed distribution of vortices in superfluid. The idea is simple, namely we put the vortices onto the homogeneous and isotropic equilibrium configuration $\Phi_{eq}(z)$ background, which can be easily obtained by numerically solving the ordinary differential equations for scalar field $\Phi_{eq}(z)$ and time component of gauge field $A_t(z)$. The resultant $\Phi_{eq}(z)$ is plotted in Fig.\ref{fig9} for $\mu=2$.

For the single vortex with winding number $\omega$, one can take the bulk configuration as
\begin{equation}
\Phi_\omega=g(r)e^{i\omega\theta}\Phi_{eq}(z),
\end{equation}
where ($r,\theta$) are polar coordinates. We require that the initial profile function $g(r)\rightarrow{1}$ for $r\rightarrow\infty$, and $g(r)\rightarrow{r^{\omega}}$ for $r\rightarrow{0}$ to guarantee the smoothness of $\Phi_\omega$ at the core of vortex. The concrete form of the initial profile function $g(r)$ will not affect the intrinsic vortex dynamics in superfluid, because the unwanted modes will be dissipated rapidly into the black hole. Then the bulk configuration for a variety of vortices can be readily constructed by multiplying $\Phi_{eq}(z)$ simultaneously with $g(r-r_i)e^{i\omega_i\theta_i}$. Regarding the lattice of vortices, one is generically required to further multiply the above bulk configuration with a random phase $e^{i\chi(x,y)}$ to make vortices get moved. For the lattice of vortices in our periodic box, we take $\chi(x,y)=Re\gamma\sum_{k_{x}=-n\Lambda}^{n\Lambda}\sum_{k_{y}=-n\Lambda}^{n\Lambda}\alpha(\boldsymbol{k})e^{i\boldsymbol{k}\cdot\boldsymbol{x}}$, where $\gamma$ is a small constant, $n$ is a small integer with $\Lambda=\frac{2\pi}{100}$, and $\alpha(\boldsymbol{k})$ is a set of $O(1)$ random complex coefficients.

\bibliographystyle{spphys}
\bibliography{thereference}
\end{document}